\documentclass[journal]{IEEEtran}                   
\usepackage{epsfig}
\usepackage{latexsym} 
\usepackage{amssymb} 
\usepackage{cite}
\usepackage{graphicx}
\usepackage{subfig}

\begin{document}
 
\title{Self quenching  streamer mode  in quenching gases  initiated by
alpha particles}

\author{Yu.I.~Davydov  and R.~Openshaw\thanks{Yu.I.~Davydov  is with
JINR, Dubna,  Russia.}\thanks{R.~Openshaw is  with TRIUMF, Vancouver,
BC, Canada.}}

\maketitle

\thispagestyle{empty}

\begin{abstract}

The transition from a proportional  to a self quenching streamer mode,
as a function of track length  and angle, was investigated in a single
wire chamber  filled with either  pure DME or isobutane.   The chamber
was irradiated  with $^{241}$Am alpha particles.   An investigation of
multistreamer  events  in DME  due  to  alpha  particles entering  the
chamber at  20$^\circ$ with  track length 4~mm  gave an estimate  of a
dead zone, defined as the product  of dead length and dead time, to be
less than  0.1~$\mu$s$\cdot$cm.  This value  is 3 orders  of magnitude
less  than  those observed  by  other  groups  for noble  gases  based
mixtures.  No  second streamers were observed with  pure isobutane for
similar tracks.

\end{abstract}

\section{Introduction}

The  TWIST experiment~\cite{TWIST} at  TRIUMF utilizes  drift chambers
filled with pure  dimethyl either (DME)~\cite{Davydov_1,Henderson} and
a  low  pressure  time  expansion  chamber  system  filled  with  pure
\mbox{iso-C$_4$H$_{10}$}~\cite{Hu}.    The  need  to   understand  the
behaviour of the chambers motivated  a study of the gas parameters and
wire-chamber operation at  different environmental conditions and high
voltages  up  to  high  current  mode.   In particular  it  led  to  a
systematic  study  of a  self  quenching  streamer  (sqs), or  limited
streamer mode in these quenching gases.

The investigation of a sqs mode  has a decades long history. This kind
of   discharge    was   observed   for   the   first   time   in   the
1970s~\cite{Charpak,Brehin}  and  has   been  extensively  studied  by
different  groups~\cite{Alekseev_1,Battistoni,Atac,Iwatani}.   A
few models were proposed to explain the sqs discharge (see for example
\cite{Alekseev_1,Atac,Zhang,Taylor,Zalikhanov}).      However,     the
discharge is  not yet fully understood.   So it is  still important to
continue the study of this type of discharge.

Quenching gases  effectively absorb  photons emitted from  the excited
atoms in  the gas mixtures and  thus limit the development  of the sqs
discharges. In pure quenching  gases, no transitions from proportional
to  sqs mode, due  to x-ray  and beta  particle ionization,  have been
observed.  However, observations of sqs signals in some pure quenching
gases  due  to  alpha  particles  have  been  reported  in  number  of
papers~(see for example \cite{Koori_1,Nohtomi_1,Nohtomi_2}). Quenching
gases do  not emit photons able  to ionize their  own atoms.  However,
the  first  papers reporting  transition  to  sqs  mode due  to  alpha
particle ionization tried to  explain the transition from proportional
to  sqs  mode by  emitted  high  energy photons~\cite{Koori_1}.   More
recently, a  model taking into account only  ionization electrons from
the        alpha         particle        tracks        has        been
proposed~\cite{Nohtomi_1,Nohtomi_2}.

Earlier  we reported  results  of a  study  of the  transition from  a
proportional to sqs  mode in single wire chambers  with different wire
diameters   and  filled   with  pure   quenching  gases   of   DME  or
\mbox{iso-C$_4$H$_{10}$}~\cite{Davydov_2}.  DME and isobutane are well
known  as excellent  quenchers.  Chambers  were irradiated  with alpha
particles from  a $^{148}$Gd  source with an  energy of  3.18~MeV.  We
demonstrated     that     chambers     filled     with     DME     and
\mbox{iso-C$_4$H$_{10}$}  gases operate  in sqs  mode with  no visible
photon  contribution  at  least  at  studied  applied  voltages.   Our
conclusion agrees with a model proposed in~\cite{Nohtomi_1,Nohtomi_2}.
Observation of double  streamer signals in DME filled  chambers due to
inclined tracks  from alpha particles  has motivated further  study of
these signals.  Alpha particles with  higher energy allow the study of
multi streamer events  as a function of track  length and angle.  This
paper  continues the study  with alpha  particles from  the $^{241}$Am
source.

\section{Experimental results and discussion}

\subsection{Experimental setup and procedures}

A  single wire chamber  with square  12x12~mm$^{2}$ cross  section and
50~$\mu$m diameter gold plated tungsten wire was employed to carry out
the tests.   The chamber was  made of aluminum alloy  with 6.35~$\mu$m
thick  aluminized mylar  windows on  two sides.   The wire  length was
approximately 20~cm. Positive high voltage was applied to the wire.

Instrument grade (0.995  purity) DME and isobutane gases  were used in
the tests.  All tests were made at atmospheric pressure.

\begin{figure*}[ht]
\centerline{\subfloat{\includegraphics[width=16cm]{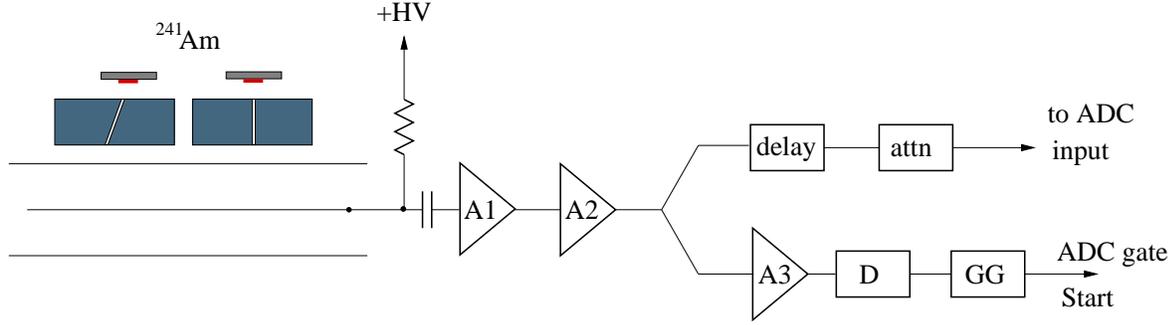}%
} 
\hfil }
\caption{The block diagram of the test setup}
\label{setup}
\end{figure*}

A  $^{241}$Am alpha  source was  used for  all tests.   The collimated
alpha source was placed directly over  the wire in such a way that the
wire and  alpha particle tracks  are near co-planar.   Alpha particles
entered the chamber either  normally or at 20$^\circ$ as schematically
presented in  Fig.~\ref{setup}.  Track  lengths inside of  the chamber
cell  were selected  by  moving  an $^{241}$Am  source  away from  and
towards the  chamber.  The ion range-energy  code SRIM~\cite{SRIM} was
used to  estimate the  alpha particle track  ranges and  energy losses
inside of the chamber cell.

\begin{figure}
\centering
\includegraphics[width=8.5cm]{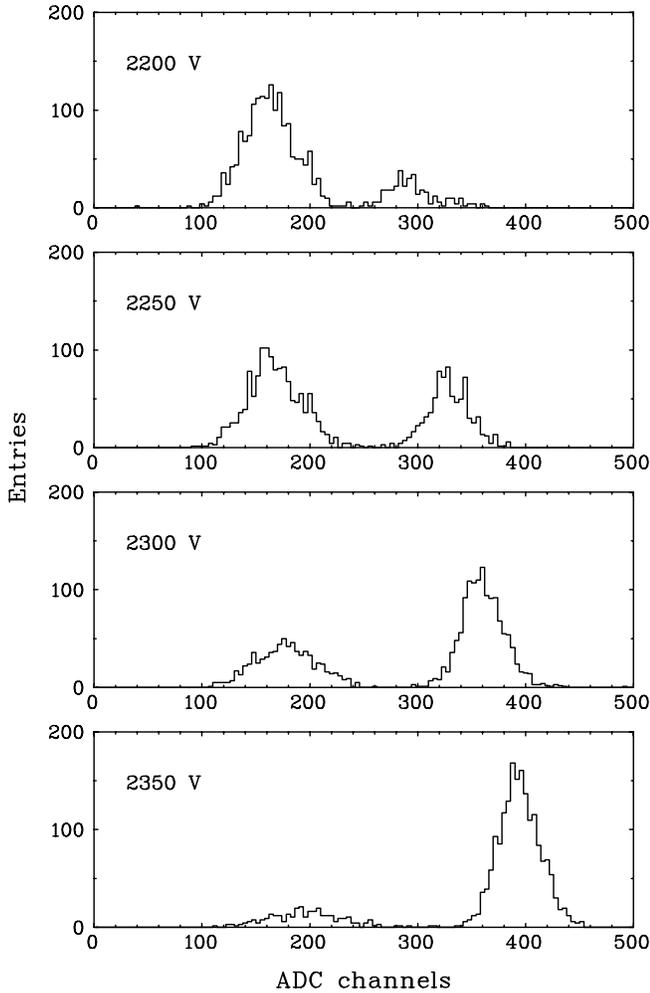}
\caption{Measured  charge   spectra  in  the   transition  region  for
 different applied voltages in  pure DME filled chamber with 50~$\mu$m
 diameter anode wire.  Chamber irradiated with normally incident alpha
 particles  from the  $^{241}$Am with  4~mm track  lengths  inside the
 chamber. Events from the proportional  peak (left) move up to the sqs
 peak (right) upon high voltage increase}
\label{peaks}
\end{figure}
 
All tests  were done in a self-triggered  mode.  Current preamplifiers
with different gains  (A1 and A2, see Fig.~\ref{setup})  were used for
the  proportional  and  sqs   modes  and  in  the  transition  region.
Amplified signals  were split and one  part was sent to  the ADC input
through the delay line (delay) and attenuator (attn).  The second part
of the  split signal was sent to  the extra amplifier A3  (used at low
voltages  at  proportional  mode  only)  and  discriminator  D.   Gate
generator (GG) provided the  ``Start'' signal for the data acquisition
system and an ADC gate.  A LeCroy 2249W model ADC was employed for the
tests.  The ADC gate signals had a duration of 2.5~$\mu$s .

Charge  spectra were  taken for  different applied  voltages.  Single
peaks related  to a proportional mode  were observed in  both gases at
low  applied  voltages. Increasing  the  high  voltage  resulted in  a
transition   from  proportional   to  sqs   mode  in   both   DME  and
\mbox{iso-C$_4$H$_{10}$}  gases.  Both  proportional  and sqs  signals
co-exist in  the transition region.  Charge spectra  in the transition
region  have two  peaks  due  to proportional  and  sqs signals.   The
fraction  of  events in  the  sqs  peak  increases while  proportional
signals  moved  up to  the  sqs peak  with  high  voltage increase  as
demonstrated in  Fig.\ref{peaks} for DME  gas at 2200, 2250,  2300 and
2350~V  applied voltages.   The chamber  was irradiated  with normally
incident alpha  particles with 4~mm  track lengths inside  the chamber
cell.

For all measurements,  mean charge values in the  proportional and sqs
peaks were found by fitting the corresponding charge spectra.

\subsection{Results and discussion}

\begin{figure}[h]
\centering \includegraphics[width=8.5cm]{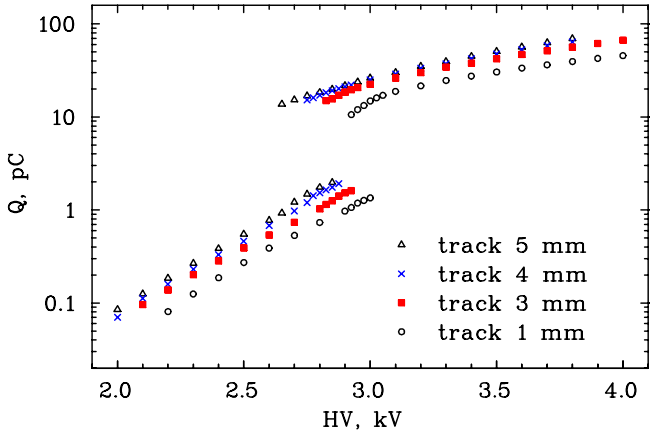}
\caption{Measured  charge  as  a  function  of high  voltage  in  pure
\mbox{iso-C$_4$H$_{10}$} filled chamber  with 50~$\mu$m diameter anode
wire.  Chamber irradiated with  normally incident alpha particles from
$^{241}$Am.  Track lengths inside the chamber are 1, 3, 4 and 5~mm.}
\label{Q_iso_0degree}
\end{figure}

Figure~\ref{Q_iso_0degree} shows the collected charge as a function of
high  voltage in pure  \mbox{iso-C$_4$H$_{10}$} filled  chamber. Alpha
particles entered  the chamber normally.  Track lengths  inside of the
chamber  volume  were   1,  3,  4  and  5~mm.    The  transition  from
proportional to  sqs mode starts  earlier on the longer  tracks, where
initial and total charges are bigger. It is interesting to notice that
the difference  in the sqs mode  between 1~mm and 3~mm  long tracks is
much bigger  than that between  3 and  4 and 5~mm.   In the case  of a
short  track, it  seems the  sqs  signals are  not developed  totally.
Small differences  between the sqs signals  from 3, 4  and 5~mm tracks
could indicate  that only part  of ionization electrons from  the long
tracks are contributing to the formation of the sqs signals.

\begin{figure}
\centering \includegraphics[width=8.5cm]{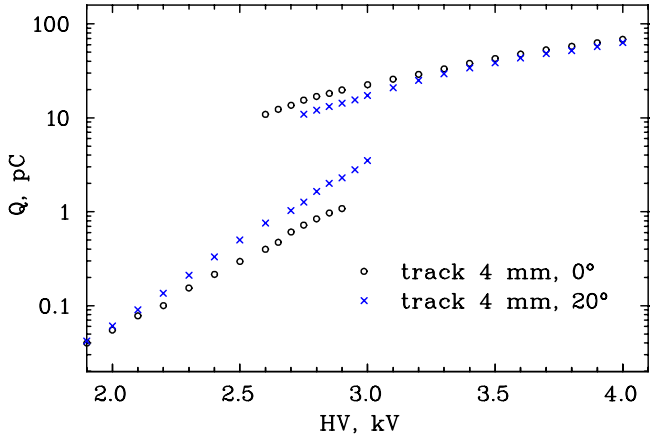}
\caption{Measured  charge  as  a  function  of high  voltage  in  pure
\mbox{iso-C$_4$H$_{10}$}  filled  chamber.   Chamber  irradiated  with
alpha  particles from  the $^{241}$Am  entering at  20$^\circ$.  Track
lengths inside of the chamber are 2 and 4~mm.}
\label{Q_iso_20degree}
\end{figure}

Differences in the  collected charges in pure \mbox{iso-C$_4$H$_{10}$}
due  to 4~mm  long tracks  from alpha  particles entering  the chamber
volume     normally    and    at     20$^\circ$    are     shown    in
Fig.\ref{Q_iso_20degree}.  Projection of a such 4~mm inclined track on
the wire length is about  1.37~mm. Initial ionizations are similar for
both cases.   However, the space charge influence  in the proportional
mode  is higher  for  normal tracks,  resulting  in smaller  collected
charge compared with that for  inclined tracks.  Transition to the sqs
mode happens at  lower voltage for normal tracks,  and their collected
charge  at the  beginning of  the sqs  mode is  bigger.   Further high
voltage  increase resulted  in the  equalization of  collected charges
from both types of tracks.   Again, this could indicate that only part
of the ionization electrons from the track are contributing to the sqs
signal.
 
\begin{figure}[h]
\centering \includegraphics[width=8.2cm]{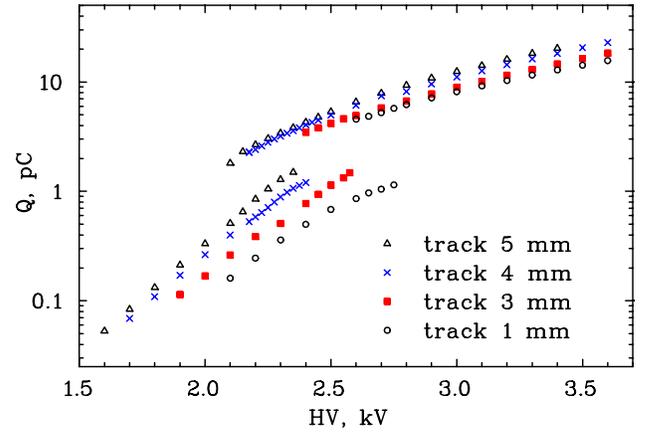}
\caption{Measured charge  as a  function of high  voltage in  pure DME
filled  chamber.   Chamber  irradiated  with normally  incident  alpha
particles from $^{241}$Am.  Track lengths inside the chamber are 1, 3,
4 and 5~mm.}
\label{Q_dme_0degree}
\end{figure}

DME gas is an even better quencher than pure \mbox{iso-C$_4$H$_{10}$}.
Collected  charge  as a  function  of high  voltage  for  pure DME  is
depicted in Fig.\ref{Q_dme_0degree}.   The chamber was irradiated with
normally  incident alpha  particles, track  lengths  are 1,  3, 4  and
5~mm. As in case of pure \mbox{iso-C$_4$H$_{10}$}, the transition from
proportional to sqs  mode starts earlier on the  longer tracks. In the
sqs mode, the  jump between 1 and  3~mm or 4 and 5~mm  is smaller than
that between 3 and 4~mm.  This  could indicate that sqs signals from 1
and 3~mm long tracks are not fully developed.
 
\begin{figure}
\vspace{0.4cm}                                               
\centering
\includegraphics[width=8.5cm]{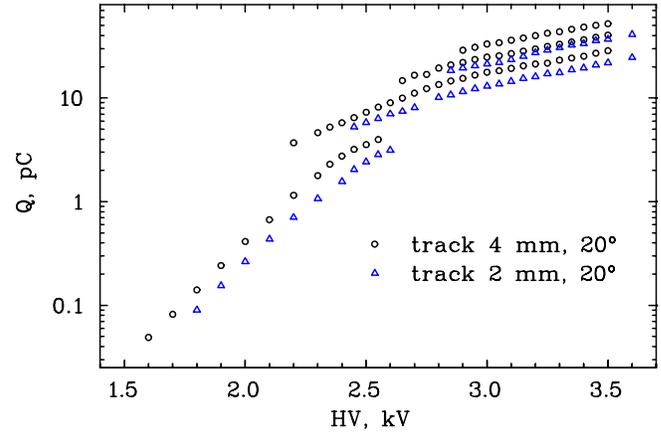}
\caption{Measured charge  as a  function of high  voltage in  pure DME
filled  chamber.  Chamber  irradiated  with alpha  particles from  the
$^{241}$Am entering at 20$^\circ$. Track lengths inside of the chamber
are 2 and 4~mm.}
\label{Q_dme_20degree}
\end{figure}

In DME, for alpha particles  entering the chamber volume at 20$^\circ$
the   situation  differs   from  that   for  isobutane.    Similar  to
\mbox{iso-C$_4$H$_{10}$},  in DME for  both 2~mm  and 4~mm  long alpha
particle tracks  high voltage increase  resulted in the  appearance of
sqs signals.  Further high voltage increase resulted in the appearance
of double  sqs signals for both  track lengths. In the  case of double
sqs signals, two streamers are  separated by a few hundred nanosecond.
Even triple sqs signals appeared due to 4~mm long tracks.  As has been
demonstrated  in Ref.~\cite{Davydov_2},  these double  and  triple sqs
signals are solely due to ionization electrons from the alpha particle
tracks.   Collected charges  from 2  and 4~mm  long tracks  from alpha
particles entering  chamber at 20$^\circ$  are shown as a  function of
high voltage in Fig.\ref{Q_dme_20degree}.

\begin{figure}
\centering
\includegraphics[width=8.5cm]{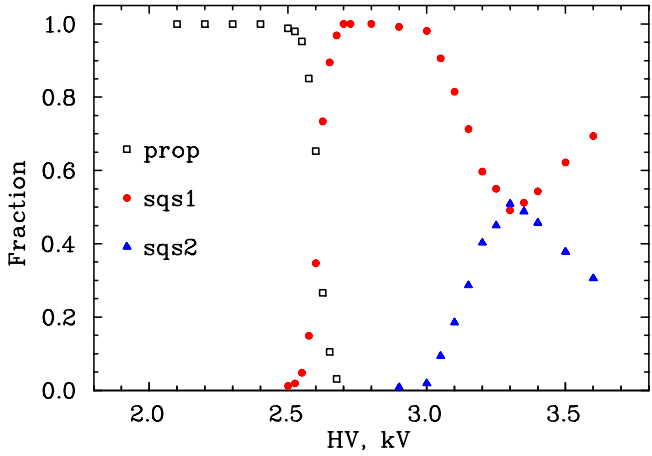}
\caption{Fractions of proportional (prop) as well as single (sqs1) and
double (sqs2) sqs signals as a  function of high voltage in a pure DME
filled  chamber.    Chamber  irradiated  with   alpha  particles  from
$^{241}$Am entering at 20$^\circ$. Track lengths inside of the chamber
volume are 2~mm.}
\label{frac_dme_2mm_20degree}
\end{figure}

\begin{figure}
\centering
\includegraphics[width=8.5cm]{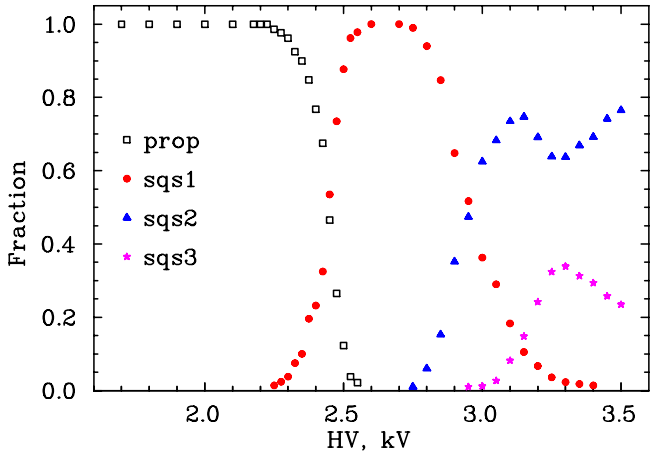}
\caption{Fractions of  proportional (prop)  as well as  single (sqs1),
double  (sqs2) and triple  (sqs3) sqs  signals as  a function  of high
voltage in a  pure DME filled chamber.  Chamber  irradiated with alpha
particles  from  $^{241}$Am  entering  at 20$^\circ$.   Track  lengths
inside of the chamber volume are 4~mm.}
\label{frac_dme_4mm_20degree}
\end{figure}

The fractions of proportional and sqs signals in DME gas as a function
of  high  voltage  for  2~mm   and  4~mm  long  tracks  are  shown  in
Fig.\ref{frac_dme_2mm_20degree}   and  Fig.\ref{frac_dme_4mm_20degree}
respectively.  In the case of 2~mm tracks, proportional signals (prop)
at first transfer to 100\%  sqs (sqs1).  Further high voltage increase
resulted  in  the  appearance  of  double sqs  signals  (sqs2),  while
decreasing the fraction of single  sqs signals. The fraction of double
sqs signals reaches a maximum,  and then decreases.  Two reasons could
be responsible for such a behaviour.  First, the high voltage increase
causes bigger avalanche  size in the first sqs  signals, which results
in an  increase of a  dead length and  dead time. The  second possible
explanation is the decrease of  the electron drift time spread and the
involvement of most  of the initial ionization electrons  in the first
sqs  avalanche. Thus, an  insufficient number  of electrons  remain to
develop the second  sqs discharge (notice that in  DME, photons do not
play any visible role in the formation of the sqs discharge).
 
Similar   transitions  occur  in   the  case   of  4~mm   long  tracks
(Fig.\ref{frac_dme_4mm_20degree}).   At  first,  proportional  signals
(prop) transfer to 100\% single sqs signals (sqs1).  Double (sqs2) and
then triple  (sqs3) sqs signals  appear with increasing  high voltage.
As  a  result, the  fraction  of  single  sqs signals  decreases,  and
eventually goes to 0.  The fraction of triple sqs signals increases to
some  maximum value,  and then  goes  down with  further high  voltage
increase.  This behaviour  could again be explained by  an increase of
dead length  and dead time  due to bigger  charge size, as well  as by
smaller drift time spread of the ionization electrons.
 
The    main     reason    for    differences     between    DME    and
\mbox{iso-C$_4$H$_{10}$}  is  the  difference  in the  electron  drift
velocities in  the two gases.   The electron drift velocities  of both
gases are  non-saturated.  At operating voltages  the reduced electric
field changes  from approximately 1~kV/cm  to 10~kV/cm within  90\% of
the drift cell  size. Over that field range the  drift velocity in DME
changes    more    than    10    times   while    the    change    for
\mbox{iso-C$_4$H$_{10}$} is less than 2 times.

\begin{figure}
\centering
\includegraphics[width=8.5cm]{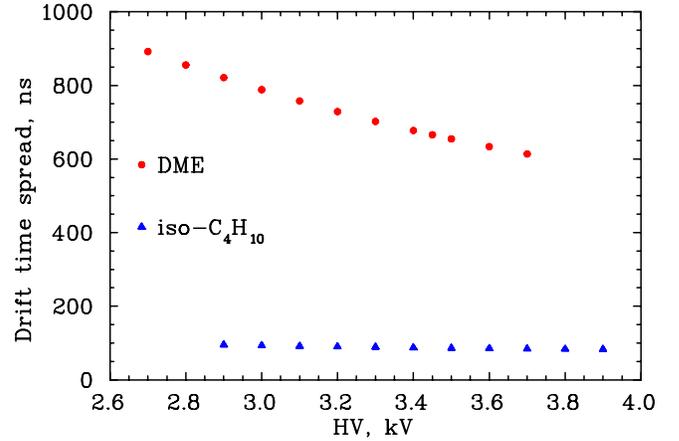}
\caption{GARFIELD  estimation   of  drift  time  spread   in  DME  and
\mbox{iso-C$_4$H$_{10}$} as a function of applied voltage.}
\label{drift_spread} 
\end{figure}

Different  drift velocities result  in essentially  different electron
drift  time spreads,  i.e. difference  in the  drift time  between the
first  and the  last  electrons  from the  alpha  particle track.   In
isobutane a drift time spread of ionization electrons from an inclined
4~mm long track in the sqs mode is approximately 80-100~ns, while that
for the same track in DME changes from approximately 1000~ns to 600~ns
with high  voltage increase.  GARFIELD~\cite{Garfield}  estimations of
the drift time spreads in sqs  mode for 4~mm inclined tracks are shown
for both gases in Fig.\ref{drift_spread} .

Such a  difference in drift time  spread could explain  the absence of
multistreamers in  pure \mbox{iso-C$_4$H$_{10}$} and  behaviour of the
multistreamer fractions  in the  case of DME  for inclined  tracks. In
fact, the  drift time spread in pure  \mbox{iso-C$_4$H$_{10}$} is less
than 100~ns, which is comparable  to a typical sqs signal width. Thus,
all electrons from  the alpha particle tracks, which  are not involved
in the streamer, are collected while  part of the wire in the vicinity
of the streamer  is insensitive to subsequent electrons.   In the case
of  DME, the  fraction  of triple  sqs  signals from  the 4~mm  tracks
increases to some  maximum value with high voltage  increase, but then
drops down.  Obviously, this  is due to  a drift time  spread decrease
with a high voltage increase.
  
\section{On the dead zone in the sqs mode}

In wire  chambers operating in sqs mode,  the part of the  wire in the
vicinity of the streamer  becomes insensitive to subsequent electrons,
i.e. the wire has a dead length $\delta$.  This part stays insensitive
during the dead  time period $\it{t_d}$.  The dead  zone $\eta$ in the
sqs mode was  defined in Ref.\cite{Alekseev_1} as the  product of dead
length and dead time: $\eta = \delta \cdot \it{t_d}$.

Typical values of the dead zone obtained by different groups for noble
gas  based   mixtures  are  within   85-400~$\mu$s$\cdot$cm  (see  for
example~\cite{Alekseev_1,Koori_2,Nohtomi}).   In  the  case  of  multi
streamer events  a dead  zone $\eta$ can  be estimated by  taking into
account dead zones for single, double, triple, etc. streamers weighted
with their fractions. Thus,
 
\begin{eqnarray}\label{eqn:Dzone}
 \eta=\sum_{i}f_i\cdot \delta_i \cdot t_{di} , \nonumber
\end{eqnarray}

\noindent  where $f_i$, $\delta_i$  and $t_{di}$  are the  fraction of
events,  dead  length  and  dead  time for  the  {\it  i}-th  streamer
multiplicity respectively.

For \mbox{iso-C$_4$H$_{10}$} one can calculate only the lower limit of
the  dead zone.   In  fact, a  4~mm  long track  at  20$^\circ$ has  a
projection on the wire length of about 1.37~mm.  The absence of double
streamers  in  isobutane suggests  that  the  dead  length for  single
streamers in  this gas  is more  than that value,  i.e. $\delta_1  > $
1.37~mm,  and  the dead  time  is more  than  the  drift time  spread,
i.e. $t_{d1} > $ 95~ns. These  numbers give a lower limit for the dead
zone   in  sqs   mode   in  \mbox{iso-C$_4$H$_{10}$}   at  the   level
$\eta>13\cdot 10^{-3}~\mu s \cdot cm$.

For multistreamer  events in  DME a dead  zone estimation can  be made
under  some simplifying  assumptions.  For  simplicity one  can assume
that single streamers have a dead length equal to the track projection
on the  wire and a  dead time equal  to the drift time  spread, double
streamers are  separated by  the length of  a track projection  on the
wire,  and triple streamers  are equidistant  on the  same projection.
Thus,  for estimation  purposes, dead  lengths for  single  and double
streamers  $\delta_1$ and  $\delta_2$ will  be taken  as equal  to the
projection  of  the track  on  the  wire,  i.e. 1.37~mm.   For  triple
streamers  a dead  length value  $\delta_3$ is  half of  that  for the
double streamers.  Dead times for single and double streamers $t_{d1}$
and  $t_{d2}$ are  taken equal  to the  drift time  spreads  for given
applied voltages. Dead time for  triple streamers $t_{d3}$ is equal to
half  of  the  drift time  spread.   It  should  be noted  that  these
simplified  assumptions underestimate contributions  to the  dead zone
from single  streamers and overestimate contributions  from double and
triple streamers.

\begin{figure}
\centering \includegraphics[width=8.5cm]{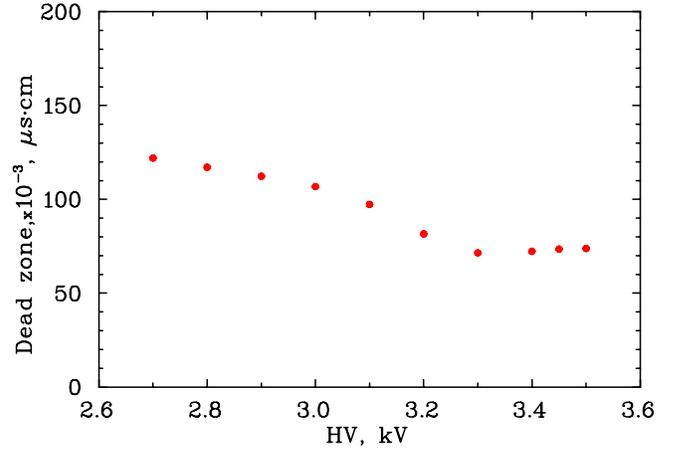}
\caption{Estimated dead zone in  the tested single wire chamber filled
with DME as a function of applied voltage.}
\label{DME_dead_zone} 
\end{figure}

Dead zones calculated under  the above mentioned assumptions with dead
time    values    taken   from    the    drift    time   spreads    in
Fig.\ref{drift_spread} and fractions of  single, double and triple sqs
signals   from    Fig.\ref{frac_dme_4mm_20degree}   are   plotted   in
Fig.\ref{DME_dead_zone}.  At  voltages corresponding to  the left-most
points of the plot, the fraction of single streamers is close to 1 and
dead  zone values  here are  well  underestimated.  The  size of  this
underestimation  decreases  as  the  fraction of  single  sqs  signals
decreases with  a high voltage  increase. At U=3.3~kV the  fraction of
single  streamers is  close  to 0  and  the dead  zone  is defined  by
contributions from  double and triple streamers.  At  this voltage the
dead   zone   has   a   minimum   value   equal   to   $\eta   \approx
0.072~\mu$s$\cdot$cm  and again  increases with  further  high voltage
increase.   Figure~\ref{frac_dme_4mm_20degree}  demonstrates that  the
fraction of  triple sqs signals has  a maximum at  U=3.3~kV.  Thus, in
the   case   of   DME,   both   Fig.~\ref{frac_dme_4mm_20degree}   and
Fig.~\ref{DME_dead_zone}  confirm that  the  dead zone  has a  minimum
value  at 3.3~kV  and increases  with  a high  voltage increase  above
3.3~kV.

The obtained value for the dead zone in pure DME is more than 3 orders
of magnitude  less than those reported  by other groups  for noble gas
based mixtures~\cite{Alekseev_1,Koori_2,Nohtomi}.  The main reason for
such  a difference is  the large  amount of  photons emitted  in those
mixtures.   These photons  produce  many secondary  avalanches in  the
vicinity of the original streamer. Thus,  in case of a noble gas based
mixture a pure  sqs signal is accompanied by  the secondary avalanches
spread  along  the wire.   These  secondary  avalanches increase  dead
length and dead time around  the original sqs signal.  We believe that
dead  zone values  reported for  noble gas  based mixtures  are mostly
defined by  these additional  avalanches and do  not reflect  the real
dead zone values for the sqs signals themselves.

\section{Conclusion}

Quenching  gases  \mbox{iso-C$_4$H$_{10}$}  and DME  both  demonstrate
100\% transition from  proportional to sqs mode due  to alpha particle
ionization.

Double and triple sqs signals appear  in the DME filled chamber due to
inclined tracks  from alpha particles,  as was demonstrated  for alpha
particles  entering  the  chamber   cell  at  20$^\circ$.   No  second
streamers  were  observed   for  the  same  type  of   tracks  in  the
\mbox{iso-C$_4$H$_{10}$}   filled    chamber.    Differences   between
\mbox{iso-C$_4$H$_{10}$} and DME are due to different drift velocities
and electron drift time spreads.

The dead  zone in a DME  filled chamber was estimated  from double and
triple sqs  signals due to inclined  tracks. The dead  zone depends on
the  applied  high  voltage  and  was found  to  be  about  0.072~$\mu
s\cdot$cm at the  minimum point.  This value is more  than 3 orders of
magnitude less than those obtained by other groups for noble gas based
mixtures.   These  reported  dead  zone  values for  noble  gas  based
mixtures are  defined by secondary  avalanches due to  emitted photons
and do not reflect real dead zone values for ``pure'' sqs signals.

\end{document}